\documentclass[aps,pra,twocolumn,showpacs,groupedaddress]{revtex4}
\usepackage{graphicx}
\usepackage{CJK}
\newcommand{\Tr}{\text{Tr}}
\newcommand{\ket}[1]{|#1\rangle}
\newcommand{\bra}[1]{\langle#1|}

\newcommand{\sz}{\sigma^z}
\newcommand{\sx}{\sigma^x}
\newcommand{\sy}{\sigma^y}
\newcommand{\sigmam}{\sigma^-}
\newcommand{\sigmap}{\sigma^+}
\newcommand{\inner}[2]{\langle #1|#2\rangle} 
\newcommand{\ext}[2]{|#1\rangle\langle#2|}   
\newcommand{\ad}{a^{\dag}}
\newcommand{\p}[1]{\frac{\partial}{\partial t}#1} 
\newcommand{\dt}[1]{\frac{d}{d t}#1}

\newcommand{\0}{\ket{0}}
\newcommand{\1}{\ket{1}}

\newcommand{\PRA}[3] {Phys. Rev. A {\bf #1}, #2
(#3)}

\newcommand{\PRD}[3] {Phys. Rev. D {\bf #1}, #2
(#3)}

\newcommand{\PRL}[3] {Phys. Rev. Lett. {\bf #1}, #2
(#3)}

\newcommand{\JPB}[3] {J. Phys. B {\bf #1}, #2 (#3)}
\newcommand{\PLA}[3] {Phys. Lett. A {\bf #1}, #2 (#3)}

\newcommand{\PS}[3] {Phys. Scr. {\bf #1} #2
(#3)}

\newcommand{\JCP}[3] {J. Chem. Phys. {\bf #1}, #2
(#3)}

\usepackage[latin1]{inputenc}
\begin{document}
\begin{CJK*}{GB}{gbsn}

\title{Born-Oppenheimer approximation  for open quantum systems within
the quantum trajectory
approach}
\author{X. L. Huang(»ÆÏþÀí) }
\email{ghost820521@163.com}
\author{S. L. Wu(ÎäËÉÁÖ)}
\author{L. C. Wang(ÍõÁÖ³É)}
\author{X. X. Yi(ÒÂÑ§Ï²)}
\email{yixx@dlut.edu.cn}
\affiliation{School of physics and optoelectronic technology,\\
Dalian University of Technology, Dalian 116024 China}

\date{\today}

\begin{abstract}
Based on the quantum trajectory approach, we extend the
Born-Oppenheimer (BO) approximation from closed quantum system  to
open quantum system,  where the open quantum system  is described by
a master equation in Lindblad form. The BO approximation is defined
and  the validity condition is derived. We find that the dissipation
in fast variables benefits the BO approximation that is different
from the dissipation in slow variables.  A detailed comparison
between this extension and our previous approximation (that is based
on the effective Hamiltonian approach, see X. L. Huang and X. X. Yi,
\PRA{80}{032108}{2009}) is presented. Several new features and
advantages are analyzed,  which show that   the two approximations
are complementary to each other. Two examples are taken to
illustrate our method.

\end{abstract}

 \maketitle

\end{CJK*}
The adiabatic and Born-Oppenheimer (BO) approximations are  among
the oldest approaches   in  quantum mechanics
\cite{Born1930AP,Born1985ZPhys}. The adiabatic approximation tells
us that
\cite{MacKenzie2007PRA,Tong2005PRL,Tong2007PRL,Zhao2008PRA,Wu2005PRA,
MarzlinPRL}  for a time-dependent  system governed by Hamiltonian
$H(t)$, if the system is prepared in one of the eigenstate
$\ket{n(t=0)}$ of $H(t=0)$ at $t=0$, it will keep in that eigenstate
$\ket{n(t)}$ of $H(t)$ at arbitrary time $t>0$ provided  the
Hamiltonian $H(t)$ is changed slowly enough.  The Born-Oppenheimer
approximation was  first given  by Born and Oppenheimer in 1927
\cite{Born1930AP}, which can be formulated as follows
\cite{CPSun1990}: Treating  the slow variables as parameters, we
first solve the fast variables with fixed slow variables. Using
these solution we obtain an effective Hamiltonian for the slow
variables. This effective Hamiltonian contains an effective vector
potential induced by fast variables. Based on this Hamiltonian we
can obtain a wave function with the slow variables. Thus the total
wave function can be factorized into a product of two wave functions
corresponding to the fast and slow variables. This method has been
widely used in physics and quantum chemistry and becomes  a
fundamental tool in these fields
\cite{Shu2009PRA,Cederbaum2008JCP,Gong2009PRA,Yi08,Bubin2009JCP,Muskatel2009PS,
Shim2009PLA,ABramov2009JPB,Forre2009PRL,Ruban2008RPP,Hua2009PRA,
Tiemann2009PRA,Leth2009PRL}.

Due to the unavoidable coupling of quantum  systems to its
environments, most quantum  systems are  open and
dissipative\cite{weiss93}. The dynamics of open quantum system can
be described by the master equation\cite{quantumnoisy,Breuerbook}.
It is then natural to ask: How to extend these approximations  from
closed system to open  system? The adiabatic approximation has been
extended to open systems in different ways,  including the Jordan
blocks method in Liouville space
\cite{Sarandy2005PRL,Sarandy2005PRA}, the effective Hamiltonian
approach \cite{Huang2008PRA,Tong2008PLA,Yi2007JPB} and in weak
dissipation limit \cite{Thunstrom2005PRA}. Although  the
approximation based on the effective Hamiltonian approach  is
equivalent to the Jordan blocks method\cite{Yi2007JPB}, the
effective Hamiltonian approach has advantages  that the extension is
straightforward and the effective Hamiltonian is easy to  give. The
BO approximation has been extended based on effective Hamiltonian
approach in Ref.\cite{Huang2009PRA}.  In this paper, we shall extend
the BO approximation in a different way, which is based on the
quantum trajectory approach. Compared with our previous work, this
extension exhibits the following  new features and advantages. (1)
We do not need to extent the Hilbert space. This will save the CPU
(computing) time and memory. (2) All eigenstates of the effective
Hamiltonian in the method are physical states. (3) The eigenstates
are easy to obtain. (4)It is more  accurate to treat the jump terms
in the master equation than that in our previous method.

Consider a quantum system with two types of variables, a slow one
$\vec{X}$ and a fast one $\vec{Y}$. Then we can divide the total
Hamiltonian of the system $H$ into two parts
\begin{eqnarray}
H=H_s(\vec{X})+H_f(\vec{X},\vec{Y}),
\end{eqnarray}
where $H_s(\vec{X})$ only contains the slow variables $\vec{X}$. The
two types of degrees of freedoms couples together through
$H_f(\vec{X},\vec{Y})$. We start considering dissipation in the fast
variables first. The case of decoherence in the slow variables will
be discussed later. Assuming the dissipation is in the Lindblad
form, the dynamics  for such a system can be described  by
\begin{eqnarray} \p
\rho=-\frac{i}{\hbar}[H,\rho]+\mathcal{L}\rho,\label{masterequation}
\end{eqnarray}
where the first term on the right hand side represents an unitary
evolution while the second term denotes the dissipation. Here we
assume the dissipative term can be arranged into the Lindblad form
as
\begin{eqnarray}
\mathcal{L}\rho=\frac12\sum_k(2L_k\rho L^{\dag}_k-\rho
L^{\dag}_kL_k-L^{\dag}_kL_k\rho),
\end{eqnarray}
where $L_k=L_k(\vec{Y})$ is the Lindblad operator relevant to the
fast variables $\vec{Y}$. $L_k\rho L^{\dag}_k$ denotes the jump
term. Within the frame of quantum trajectory approach, for an
initial state $\ket{\phi(t_0)}$, one can write the state  after an
infinitesimal time $dt$ as
\begin{eqnarray}
\rho(t_0+dt)=\left(1-\sum_kdp_k\right)\ext{\phi_0}{\phi_0}+
\sum_kdp_k\ext{\phi_k}{\phi_k},
\end{eqnarray}
where $dp_k=\bra{\phi(t_0)}L_k^{\dag}L_k\ket{\phi(t_0)}dt$ and the
new states are defined by
\begin{eqnarray}
&&|\phi_0\rangle = \frac{(1-i H_{\text{eff}} dt/\hbar)
|\phi(t_0)\rangle} {\sqrt{1-\sum_{k} dp_{k}}}, \nonumber\\
&&|\phi_{k}\rangle = \frac{L_{k} |\phi(t_0)\rangle}{||L_{k}
|\phi(t_0)\rangle||}, \label{newstates}
\end{eqnarray}
with the non-Hermitian effective Hamiltonian defined by
$H_{\text{eff}}=H-\frac{i}2\hbar\sum_kL_k^{\dag}L_k$. Under this
description, the system will jump into the state $\ket{\phi_k}$ with
probability $dp_k$, and evolve according to non-Hermitian effective
Hamiltonian $H_{\text{eff}}$ with probability $1-\sum_kdp_k$. This
unraveling is the so-called Monte Carlo wave function method
\cite{Dalibard1992PRL,Carmichaelbook,Molmer1993JOSAB}. The
difficulty here is that the non-Hermitian Hamiltonian
$H_{\text{eff}}$ contains two types of variables, we will solve this
problem by applying the BO approximation in the no-jump trajectory
only. For the non-jump evolution $\ket{\phi_0}$, the time evolution
is given by
\begin{eqnarray}
i\hbar\dt\ket{\Psi(t)}=H_{\text{eff}}\ket{\Psi(t)}.\label{Shrodingerlike}
\end{eqnarray}
Our aim is to solve this equation with the help of BO approximation.
To this end, we first rewrite $H_{\text{eff}}$ as
$H_{\text{eff}}=H_s(\vec{X})+H_f'(\vec{X},\vec{Y})$ with
$H_f'(\vec{X},\vec{Y})=H_f(\vec{X},\vec{Y})-\frac{i}2\sum_kL_k^{\dag}L_k$.
Obviously, $H_f'(\vec{X},\vec{Y})$ is not Hermitian that includes
all  non-Hermitian parts of $H_{\text{eff}}$. Taking the slow
variables $\vec{X}$ as parameters, we can solve the eigenstates for
$H_f'(\vec{X},\vec{Y})$. We denote its right eigenstates by
$\ket{\psi_n^R(\vec{X})}$ and the corresponding left eigenstates by
$\bra{\psi_n^L(\vec{X})}$ with complex eigenvalues $E_n(\vec{X})$.
These eigenstates satisfy the relations
$\inner{\psi_m^L}{\psi_n^R}=\delta_{mn}$ and
$\inner{\psi_n^R}{\psi_n^R}=1$ for fixed $\vec{X}$. We also restrict
our discussion to the non-degenerate case. In order to solve
Eq.(\ref{Shrodingerlike}), we  expand the eigenstate of
$H_{\text{eff}}$ in terms of $\ket{\psi_n^R(\vec{X})}$ as
\begin{eqnarray}
\ket{\Phi}=\sum_{n=1}^Nc_n\ket{\varphi_n(\vec{X})}\ket{\psi_n^R(\vec{X},\vec{Y})},\label{expand}
\end{eqnarray}
where $N$ is the dimension of the fast variables $\vec{Y}$ and $c_n
\ (n=1,2,3,...,N)$ are the expansion coefficients. Substituting
Eq.(\ref{expand}) into the eigenvalue equation
$H_{\text{eff}}\ket{\Phi}=E\ket{\Phi}$, after simple calculation, we
obtain an equation for the wavefunction of the slow variables
\begin{eqnarray}
\sum_m\bra{\psi^L_n}H_s(\vec{X})\ket{\psi_m^R}\ket{\varphi_m(\vec{X})}+
E_n(\vec{X})\ket{\varphi_n(\vec{X})}=E\ket{\varphi_n(\vec{X})}\label{sloweqn}
\end{eqnarray}
Define $H_{n,m}(\vec{X})=\bra{\psi_n^L}H_s(\vec{X})\ket{\psi_m^R}$,
we can rewrite Eq.(\ref{sloweqn}) in a matrix form as
\begin{eqnarray}
(\mathcal{H}_0+\mathcal{H}_\mathcal{P})\varphi=E\varphi,\label{matrixform}
\end{eqnarray}
where $\mathcal{H}_0,~\mathcal{H}_\mathcal{P}$ and $\varphi$ are
defined by
\begin{eqnarray*}
\begin{array}{c}
\mathcal{H}_0=\left[%
\begin{array}{cccc}
   H_{1}{+}E_1(\vec{X})&\  0  &\ \cdots  &\ 0 \\
  0 &\  H_2{+}E_2(\vec{X}) &\ \cdots   &\ 0 \\
  \vdots &\ \vdots &\ \ddots &\ \vdots\\
  0 &\ 0&\  \cdots &\ H_N{+}E_N(\vec{X})
  \end{array}%
\right],
\end{array}
\end{eqnarray*}
\begin{equation}
\begin{array}{c}
\mathcal{H}_\mathcal{P}=\left[%
\begin{array}{cccc}
  0  &\  {H}_{1,2}  &\ \cdots  &\ {H}_{1,N} \\
  {H}_{2,1} &\  0 &\ \cdots   &\ {N}_{2,N} \\
  \vdots &\ \vdots &\ \ddots &\ \vdots\\
  {H}_{N,1} &\ {H}_{N,2}&\  \cdots &\
  0\\
  \end{array}%
\right],\ \ \
\end{array}
\begin{array}{c}
\varphi=\left[
\begin{array}{c}
  \ket{\varphi_{1}}  \\
  \ket{\varphi_{2}} \\
  \vdots\\
  \ket{\varphi_{n}}\\
  \end{array}
\right].
\end{array}
\end{equation}
Here we have omitted the same subscripts for simplicity. Treating
$\mathcal{H}_\mathcal{P}$ as perturbation, we can solve
Eq.(\ref{matrixform}) by virtue of the standard time-independent
perturbation theory. The solution to the zero-order
\begin{eqnarray}
\begin{array}{c}
\widetilde{\varphi}_{1,k}^{R[0]}=\left[
\begin{array}{c}
  \ket{\varphi_{1,k}^{R[0]}}  \\
  0 \\
  \vdots\\
  0\\
  \end{array}
\right]
\end{array},\begin{array}{c}
\widetilde{\varphi}_{2,k}^{R[0]}=\left[
\begin{array}{c}
  0 \\
  \ket{\varphi_{2,k}^{R[0]}}  \nonumber\\
  \vdots\\
  0\\
  \end{array}
\right]
\end{array},\\
\cdots,\quad\cdots,\qquad\quad
\begin{array}{c}
\widetilde{\varphi}_{N,k}^{R[0]}=\left[
\begin{array}{c}
  0 \\
  0 \\
  \vdots\\
  \ket{\varphi_{N,k}^{R[0]}} \\
  \end{array}
\right]
\end{array},
\end{eqnarray}
can be obtained by the eigenvalue equation
\begin{eqnarray}
\mathcal{H}_n\ket{\varphi_{n,k}^{R[0]}}=E_{n,k}^{[0]}\ket{\varphi_{n,k}^{R[0]}},\label{slowzeroeqn}
\end{eqnarray}
where $\mathcal{H}_n=(H_n(\vec{X})+E_n(\vec{X}))$ is the zero-order
effective Hamiltonian for the slow variables. From these
zeroth-order solutions, one can obtain the higher-order correction
accordingly. The condition with which we can neglect the
higher-order correction safely is
\begin{eqnarray}
\left|\frac{\bra{\varphi_{n',k'}^{L[0]}}H_{n',n}\ket{\varphi_{n,k}^{R[0]}}}
{E_{n',k'}^{[0]}-E_{n,k}^{[0]}}\right|\ll1,~~\text{for
all}~~k',n'\neq k,n,
\end{eqnarray}
where $\bra{\varphi_{n',k'}^{L[0]}}$ is the left eigenstate of
non-Hermitian  Hamiltonian $\mathcal{H}_n$.

Next we consider the dissipation about the slow variables. The
method used in this case is very similar to the discussion given
above. In this case, the Lindblad operator is replaced by $X_k$ and
it is a function of slow variables only, i.e., $X_k=X_k(\vec{X})$.
In the non-jump trajectory, we divide the non-Hermitian Hamiltonian
as $H_{\text{eff}}=H_s'(\vec{X})+H_f(\vec{X},\vec{Y})$, where
$H_s'(\vec{X})=H_s(\vec{X})-\frac{i}2\sum_kX_k^{\dag}X_k$. The
method to find the eigenstates and eigenvalues for $H_f$ is the same
as for a closed system. These eigenstates are denoted by
$\ket{\psi_n(\vec{X})}$ with corresponding eigenvalues
$E_n(\vec{X})$. We can handle the slow variables in the same way.
The only difference is $H_{n,m}(\vec{X})$ in Eq.(\ref{sloweqn})
defined as $H_{n,m}(\vec{X})=\bra{\psi_n}
\left(H_s(\vec{X})-\frac{i}2\sum_kX_k^{\dag}X_k\right)\ket{\psi_m}$.
The zero-order effective Hamiltonian $\mathcal{H}_n$ for slow
variables in non-jump trajectory can be similarly obtained. Compared
with the closed system, this Hamiltonian  includes a non-Hermitian
(usually anti-Hermitian) correction
$\bra{\psi_n}-\frac{i}2\sum_kX_k^{\dag}X_k\ket{\psi_n}$, which comes
from the dissipation.

According to the discussion given above, we can solve the dynamics
governed by the non-Hermitian Hamiltonian $H_{\text{eff}}$ via
expanding the total state as
$\ket{\Phi}=\sum_{n,k}c_{n,k}\ket{\varphi_{n,k}^{[0]}}\ket{\psi^{(R)}_n}$.
Then the Monte Carlo simulation for Eq.(\ref{masterequation}) is the
following. We divide the total evolution time $T$ into several
steps. The interval of each step is $dt$. In each step, a random
number $\varepsilon$ which distributes in the unit interval $[0,1]$
uniformly is chosen to determine the jump process. If
$\varepsilon\leq\sum_kdp_k$, the total state jumps into the
corresponding state according to the corresponding Lindblad
operator, i.e., for $\varepsilon\leq dp_1$, it jumps to
$\ket{\phi_1}$, for $dp_1<\varepsilon\leq dp_1+dp_2$, it jumps to
$\ket{\phi_2}\cdots$. If $\varepsilon>\sum_kdp_k$, it has no-jump
process. The system evolves according to the non-Hermitian
Hamiltonian $H_{\text{eff}}$ and the BO approximation is used. This
process is repeated as many time as $n_{\text{step}}=T/dt$, and this
single evolution gives a quantum trajectory. We can recover the
final state of the system by averaging over different quantum
trajectories.


\begin{figure}
\includegraphics*[width=0.6\columnwidth, bb=0 26 256
180]{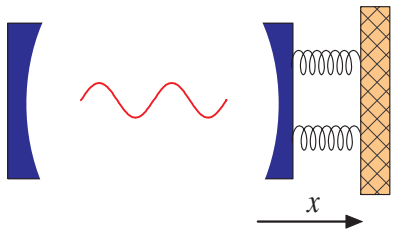}\caption{(Color online) Schematic
illustration of a Fabry-P\'{e}rot cavity with an oscillating mirror
at the right  end. }\label{FIG:example1setup}
\end{figure}

\begin{figure}
\includegraphics*[width=0.48\columnwidth, bb=88 265 490 577]{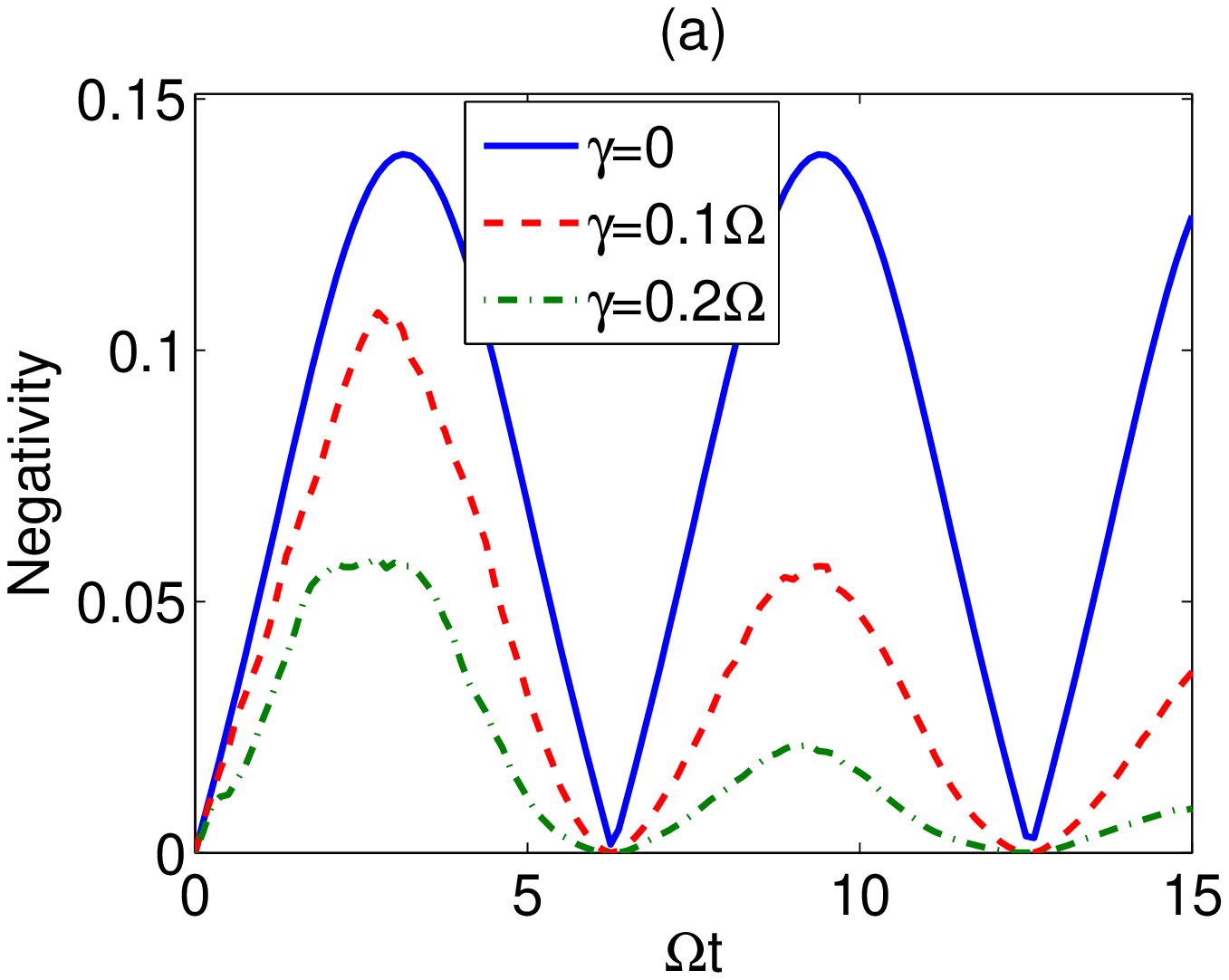}
\includegraphics*[width=0.48\columnwidth, bb=78 265 480 577]{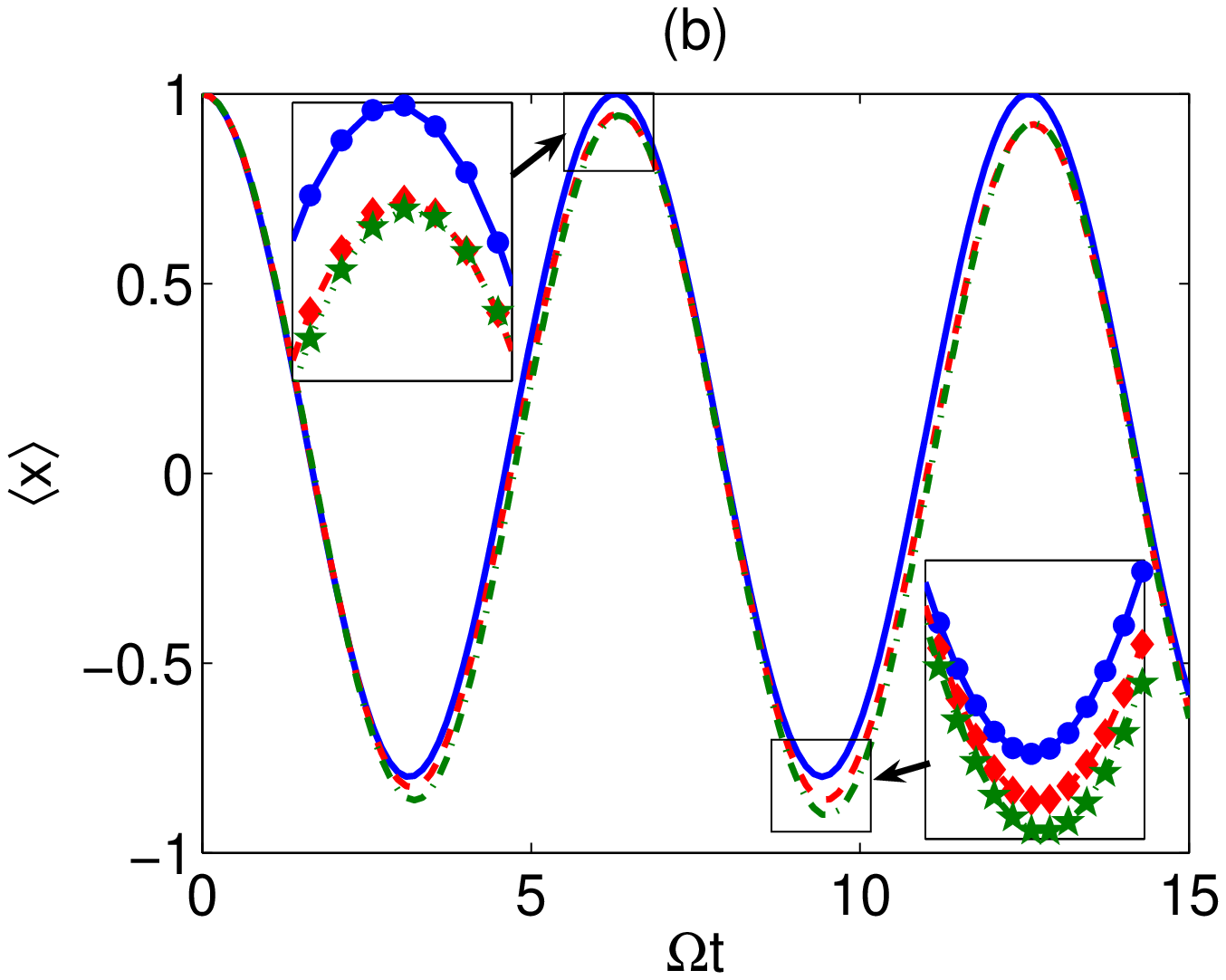}
\includegraphics*[width=0.99\columnwidth, bb=24 328 550 512]{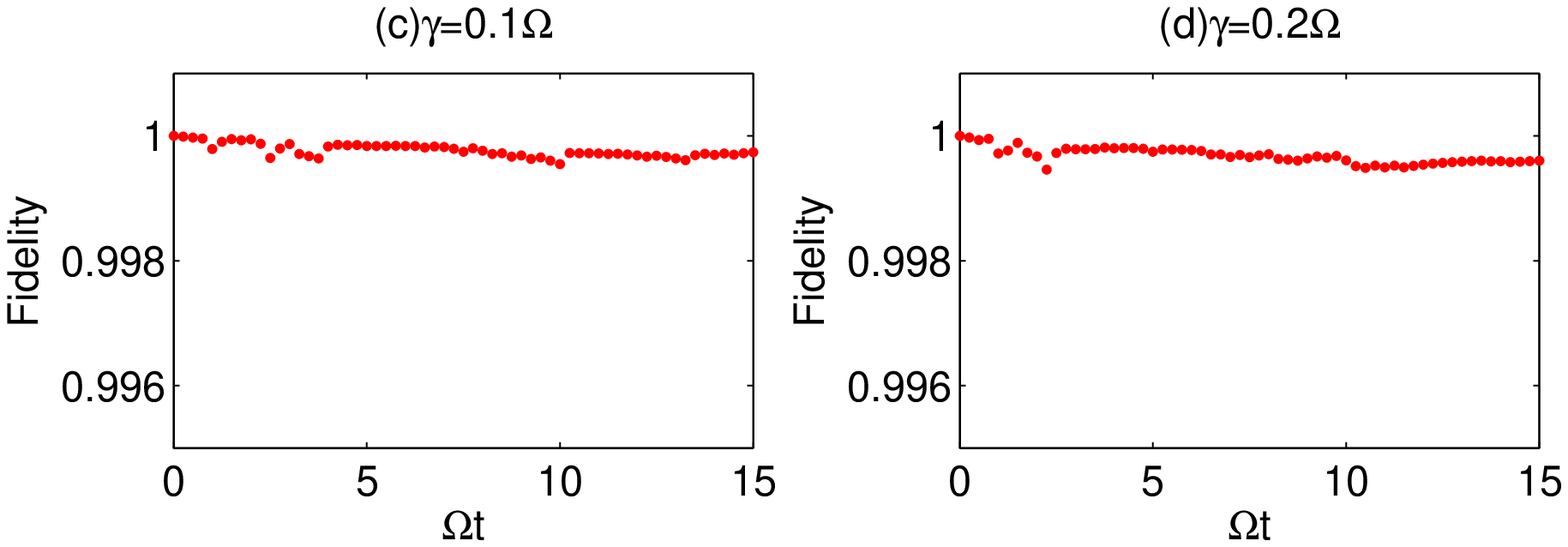}
\caption{(Color online) (a) The entanglement measured by the
negativity as a function of $\Omega t$. (b) The average value of the
coordinate $\langle x\rangle$ (in units of
$\sqrt{\frac{\hbar}{2m\Omega}}$) as a function of $\Omega t$. (c)
and (d) The fidelity between the QTA solution and numerical
simulation (Runge-Kutta method). Other parameters in the simulation
are $\omega=100\Omega$, $g=0.1\Omega$. The results are obtained by
averaging over $\mathcal{N}=150$ runs.}\label{FIG:example1_2}
\end{figure}

In the following, we shall present  two examples to illustrate our
method. After these two examples, we will give a detailed comparison
with our previous work \cite{Huang2009PRA}. First, we consider a
Fabry-P\'{e}rot(FP) cavity with an oscillating mirror at one end,
acting as a quantum-mechanical harmonic oscillator. Such system can
be described by the following Hamiltonian
\begin{eqnarray}
H=\hbar\omega\ad a-\hbar\chi\ad
ax+\frac{p^2}{2m}+\frac12m\Omega^2x^2,\label{example1H}
\end{eqnarray}
where $\omega$ is the frequency of the cavity field with the
creation and annihilation operator $a^{\dag}$ and $a$, respectively.
$m,\Omega,x$ and $p$ denote the mass, frequency, displacement, and
momentum of the oscillating mirror, respectively. $\chi=\omega/L$ is
the coupling  constant between the cavity field and the mirror. $L$
denotes the length of the cavity. Taking the cavity dissipation into
account, the Lindblad operator in this example is
$L_1=\sqrt{\gamma}a$. The non-Hermitian effective Hamiltonian for
the non-jump trajectory can be written as
\begin{eqnarray}
H_{\text{eff}}=\hbar(\omega-\frac{i}2\gamma )\ad a-\hbar g\ad
a(b+b^{\dag})+\hbar\Omega(b^{\dag}b+\frac12),
\end{eqnarray}
where $b=\sqrt{\frac{m\Omega}{2\hbar}}(x+\frac{ip}{m\Omega})$ and
$g=\chi\sqrt{\frac{\hbar}{2m\Omega}}$. Usually, the characteristic
frequency of cavity field can reach  the order of about $10^{14}$
Hz, which is much higher than the nano-mechanical resonator
frequency $10^9$ Hz achieved by current experiments
\cite{Gong2009PRA}. Under this condition, we can divide this
Hamiltonian into two parts as $H_{\text{eff}}=H_s+H_f$ with
$H_s=\hbar\Omega(b^{\dag}b+\frac12)$ and
$H_f=\hbar(\omega-\frac12i\gamma)a^{\dag}a-\hbar g
a^{\dag}a(b+b^{\dag})$. The eigenstate for the fast variables $H_f$
is $\ket{\psi_{n_a}^R}=\ket{n_a}$, where $\ket{n_a}$ is the Fock
state for the mode $a$, and corresponding left eigenstate
$\bra{\psi_{n_a}^L}=\bra{n_a}$ and eigenvalue
$E_{n_a}=\hbar(\omega-i\gamma)n_a-\hbar gn_a(b+b^{\dag})$. Putting
these into Eq.(\ref{slowzeroeqn}) and following the BO approximation
process, we obtain the Hamiltonian for the slow variables as
\begin{eqnarray}
\mathcal{H}_{n_a}{=}\hbar\Omega(b^{\dag}b{+}\frac12){-}\hbar
gn_a(b{+}b^{\dag}){+}\hbar(\omega-\frac12i\gamma)n_a.
\end{eqnarray}
This Hamiltonian can be solved by a displacement of the Fock state
\cite{Gong2009PRA} as
\begin{eqnarray*}
&&\ket{\varphi_{n_a,n_b}^R}=D(\alpha(n_a))\ket{n_b},\\
&&E_{n_a,n_b}=\hbar\Omega(n_b+\frac12)+\hbar(\omega-\frac12i\gamma)n_a-\frac{\hbar
g^2}{\Omega}n_a^2,
\end{eqnarray*}
where $D(\alpha)=e^{A^{\dag}\alpha-A\alpha^*}$ is the displacement
operator with $A=b-\alpha$, and $\alpha(n_a)=\frac{n_ag}{\Omega}$,
$\ket{n_b}$ is the Fock state for mode $b$. Note that in this model,
the off-diagonal elements of the perturbation
$\mathcal{H}_{\mathcal{P}}$ is zero, so the BO solution
$\ket{\varphi_{n_a,n_b}^R}$ for the non-jump trajectory is an exact
solution. We study the dynamics for such a Hamiltonian according to
the method given above and compare this solution to the solution
obtained by the Runge-Kutta method in Fig.\ref{FIG:example1_2}. We
choose $\ket{\Phi(0)}=\frac12(\0+\1)(\0+\1)$ as the initial state.
To make the effects of dissipation more strikingly, we choose the
parameters as $\omega=100\Omega$, $g=0.1\Omega$ in the simulation.
In Fig.\ref{FIG:example1_2}(a) we study the entanglement between the
vibration of the mirror and the cavity field. We choose negativity
\cite{Vidal2002PRA} as the measure of entanglement for mixed state.
In the simulation, the density matrix is calculated by averaging
over different runs, i.e., from the state vectors $\ket{\psi_i(t)}$
for the different trajectories,  the density matrix can be
constructed as $\rho(t) = \frac{1}{N}\sum_i^N \vert \psi_i(t)
\rangle \langle \psi_i(t) \vert$. Then we can calculate the
negativity for $\rho(t)$. We find that the Hamiltonian
Eq.(\ref{example1H}) can produce entanglement. The dissipation
decreases the entanglement gradually, and the larger the dissipation
is, the faster the entanglement decays. In
Fig.\ref{FIG:example1_2}(b) we plot the average value of the
coordinate $x$ for the oscillating mirror as a function of $\Omega
t$. From the figure we find that when the system is a closed system,
the average value of the coordinate for the mirror oscillates with
time. The dissipation  moves the curve left. Similarly, the strength
of the dissipation determines the displacement. In our simulation,
we average  our results over $\mathcal{N}=150$ runs. To check the
validity  of our method, we compare our results with the results
from the Runge-Kutta method. We use the fidelity \cite{Fidelity} as
the measure of difference between two density matrix. For mixed
state, the fidelity is defined as
$F(\rho_1,\rho_2)=\Tr\sqrt{\sqrt{\rho_1}\rho_2\sqrt{\rho_1}}$. This
fidelity reaches  1 when the two states are same. In
Figs.\ref{FIG:example1_2}(c) and (d) we plot the fidelity between
the BO solution and the Runge-Kutta method  as a function of $\Omega
t$ for different $\gamma$. It is obvious that the fidelities are
always larger than $99.9\%$ in our simulation for $\mathcal{N}=150$
trajectories \footnote{The convergence velocity of the Monte Carlo
wave function method depends on the problem considered, i.e., the
number of jump operators. Moreover for different operators, the
number of  runs $\mathcal{N}$ with which we can get an good
  result depends on the properties of the operators
themselves. For global operators, we can get good result with
relative small number of runs. This is discussed in detail in Ref.
\cite{Molmer1993JOSAB}. In our example, the  time for the
Runge-Kutta solution is as much as about $\mathcal{N}=50$ runs for
Monte Carlo wave function method. In our simulation, if
$\mathcal{N}=25$, the fidelity can reach $99\%$, if
$\mathcal{N}=50$, i.e., the time for both methods are equivalent,
the fidelity is higher than $99.5\%$. The purpose for
$\mathcal{N}=150$ in our simulation is to get a fidelity higher than
$99.9\%$. }. The error is smaller than $0.1\%$. This confirms that
our method can reproduce dissipation dynamics for open system
efficiently.

\begin{figure}
\includegraphics*[width=0.48\columnwidth, bb=88 265 490 577]{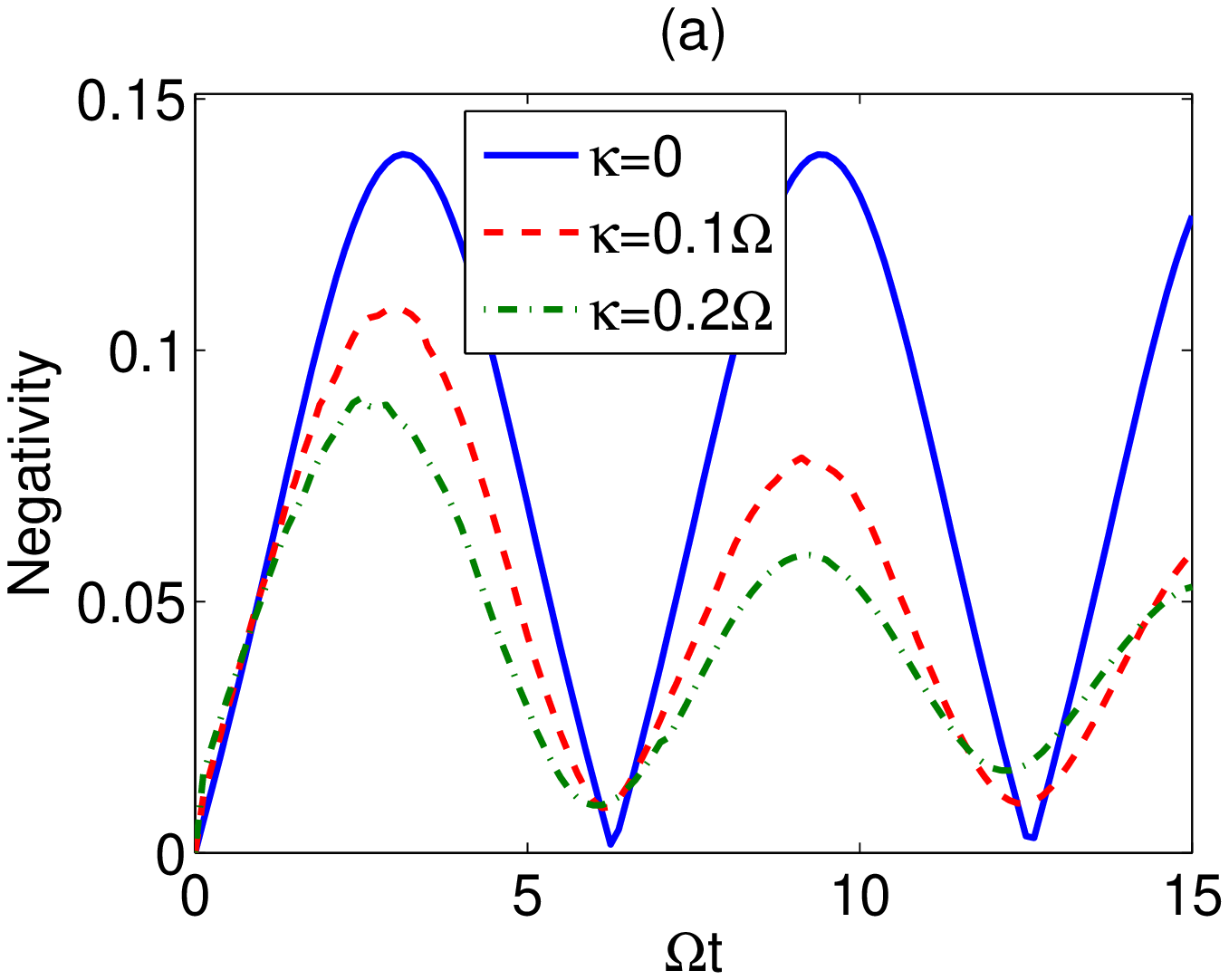}
\includegraphics*[width=0.48\columnwidth, bb=78 265 480 577]{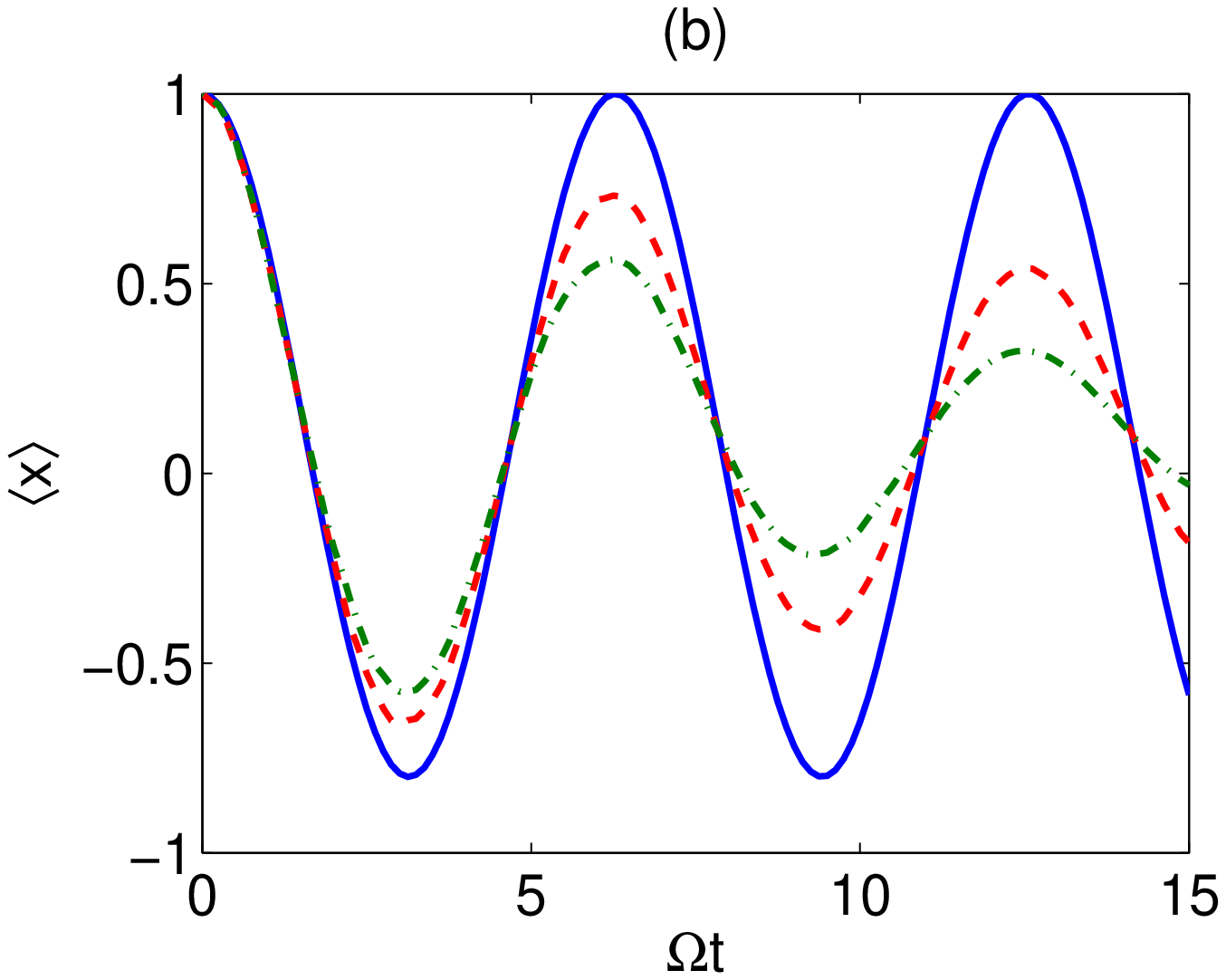}
\includegraphics*[width=0.48\columnwidth, bb=88 265 480 577]{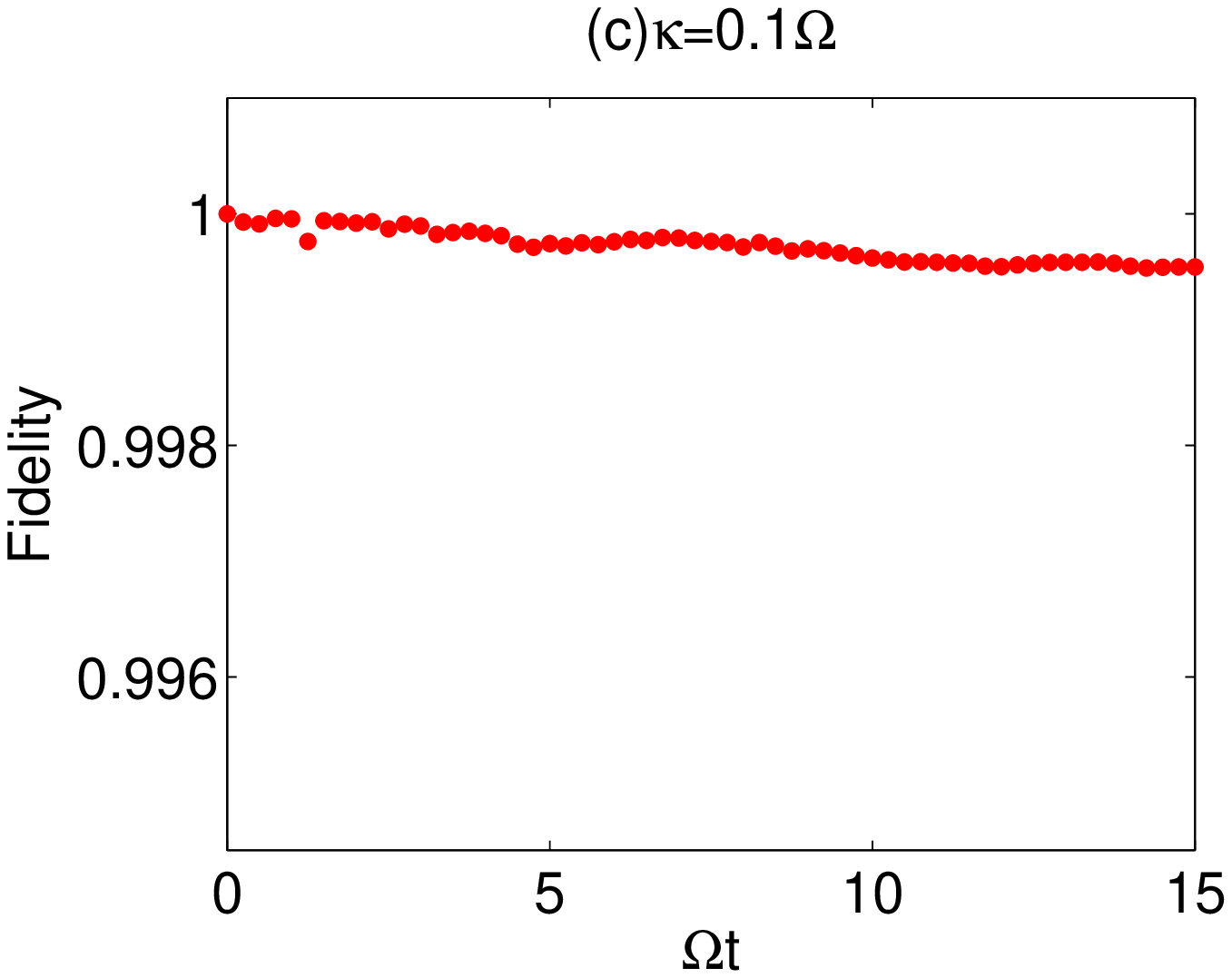}
\includegraphics*[width=0.48\columnwidth, bb=88 265 480 577]{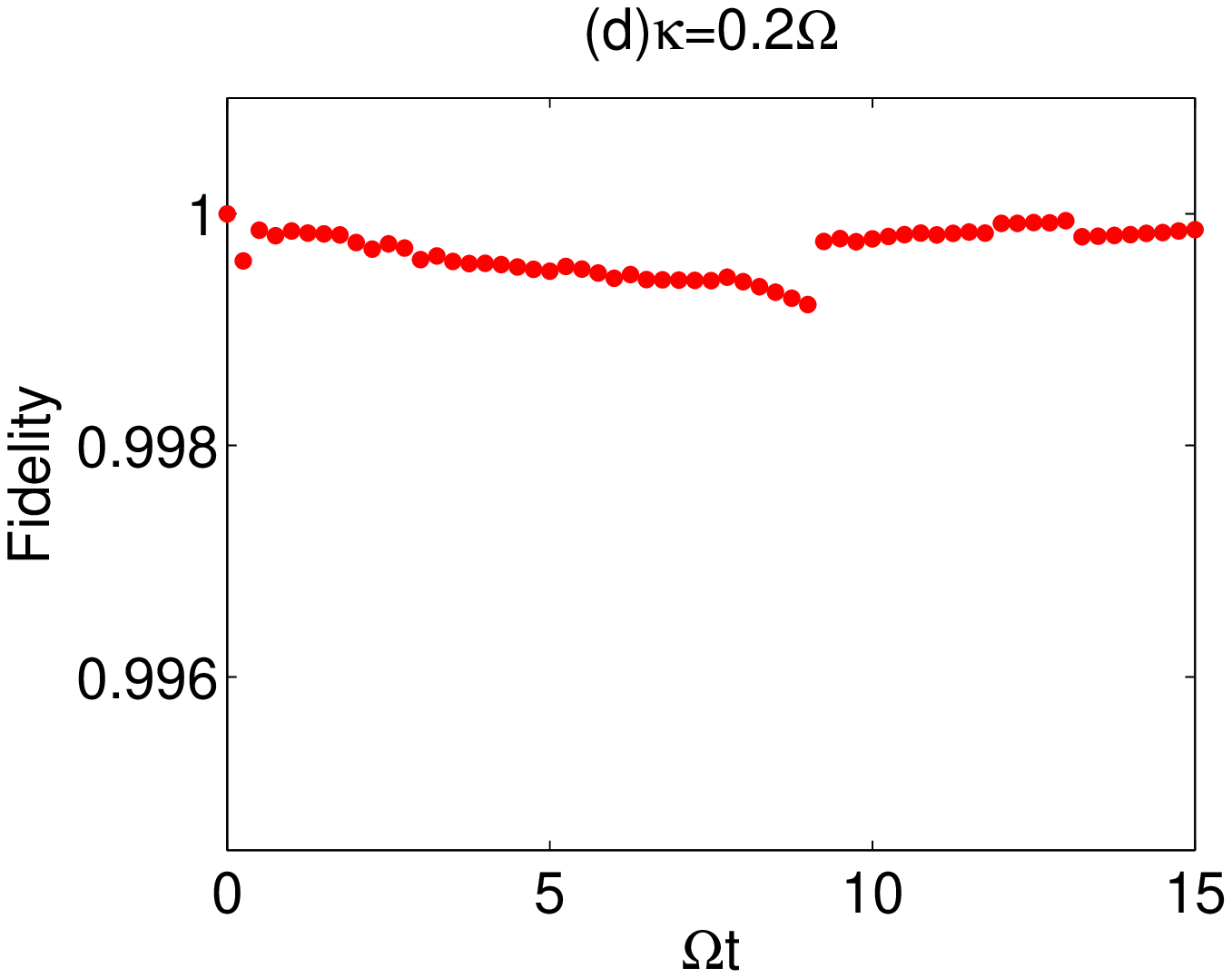}
\caption{(Color online) The same as Fig. \ref{FIG:example1_2} for
dissipation in slow variables.}\label{FIG:example1_3}
\end{figure}

Next we briefly discuss the dissipation in the slow variables for
this model. In this case, we also assume the dissipation is in the
Lindblad form, then the Lindblad operator reads $X_1=\sqrt{\kappa}b$
and the non-Hermitian effective Hamiltonian for the non-jump
trajectory is
\begin{eqnarray}
H_{\text{eff}}=\hbar\omega\ad a-\hbar g\ad a(b+b^{\dag})+
\hbar\Omega(b^{\dag}b+\frac12)-\frac12i\hbar\kappa b^{\dag}b.
\end{eqnarray}
Divide it into two parts as $H_{\text{eff}}=H_s+H_f$ with $H_s=
\hbar\Omega(b^{\dag}b+\frac12)-\frac12i\hbar\kappa b^{\dag}b$ and
$H_f=\hbar\omega\ad a-\hbar g\ad a(b+b^{\dag})$. The eigenstates for
the fast variables are $\ket{\psi}=\ket{n_a}$ with eigenvalues
$E_{n_a}=\hbar\omega n_a-\hbar g n_a(b+b^{\dag})$. With these
knowledge, we obtain the zero-order effective Hamiltonian for the
slow variables as
\begin{eqnarray}
\mathcal{H}_{n_a}=\hbar\Omega_0(b^{\dag}b+\frac12)-\hbar
gn_a(b+b^{\dag})+\hbar\omega n_a+\frac12i\hbar\kappa
\end{eqnarray}
with $\hbar\Omega_0=\hbar\Omega-\frac12i\hbar\kappa$. This
Hamiltonian can be solved in a similar way with the displacement
$\alpha_0=\frac{n_ag}{\Omega_0}$. The  calculations are similar to
the process where the dissipation in fast variables is taken into
account. The numerical results for this case is shown in Fig.
\ref{FIG:example1_3}. Two differences can be seen from the figure:
(1) When the dissipation in slow variables is taken into account,
the entanglement never disappears. (2) The amplitude for the average
value of the coordinate $\langle x\rangle$ decrease strikingly due
to the dissipation.
 We should note that in this model, the
perturbation  $\mathcal{H}_{\mathcal{P}}$ is zero in both slow and
fast variables' dissipation case. In general, this condition can not
be satisfied (see the second example and Eq.(\ref{Gammag})), then we
need to restrict  the dissipation in slow variables to be weak,
because strong dissipation in slow variables enlarge the
perturbation $\mathcal{H}_{\mathcal{P}}$, which leads  the BO
condition  to be broken down. This can be understood as follows,
large dissipation rate may accelerate the change  of slow variables,
so that it is hard to distinguish which are the fast variables.

\begin{figure}
\includegraphics*[width=0.8\columnwidth]{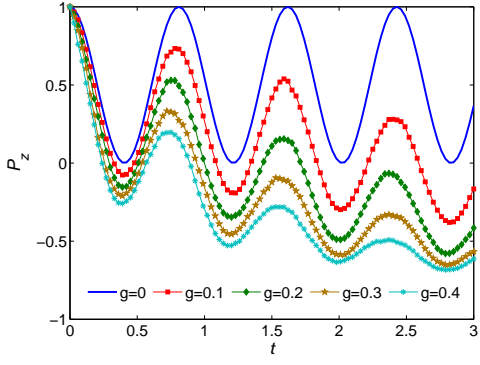}\caption{
(Color online) The polarization of the neutron along the $z$ axis as
a function of time $t$ (in units of $\pi\hbar/\mu B$) for different
dimensionless dissipation rate $g$. We have set $\theta=\pi/4$ and
initially the spin is in a state
$\ket{{+}\frac12}$.}\label{FIG:example2_1}
\end{figure}

\begin{figure}
\includegraphics*[width=0.8\columnwidth]{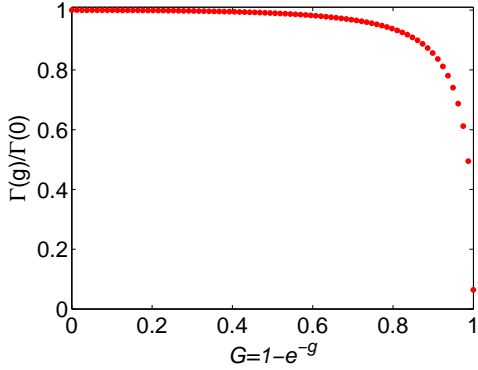}\caption{
Validity measure $\Gamma(g)$ as a function of the dimensionless
dissipation rate $g$. The results have been normalized in units of
$\Gamma(g=0)$. Other parameters in the figure are set to satisfy
$\hbar/\mu BM^2L=10^{-6}$ and $\hbar k_z/\mu
BML=2\times10^{-4}$.}\label{FIG:example2_2}
\end{figure}

 In the second example, we consider a neutron moving
in a static helical magnetic field,
\begin{eqnarray}
\vec{B}=\vec{B}(z)=B\left(\sin\theta\cos\frac{2\pi z}L,
~\sin\theta\sin\frac{2\pi z}L,~\cos\theta\right).
\end{eqnarray}
The Hamiltonian for such a system is
\begin{eqnarray}
H=H(z)=\frac{{\vec{p}}^2}{2M}+\mu\vec{B}\cdot\vec{\sigma}={H}_K+H_S,
\end{eqnarray}
where $\vec{\sigma}=(\sx,\sy,\sz)$ is the Pauli operators.  Taking
the spin relaxation into account, the Lindblad operator is
$L_2=\sqrt{\kappa}\sigmam$. Treat the coordinate as parameters, the
non-Hermitian Hamiltonian for the non-jump trajectory can be written
as
\begin{eqnarray}
H_{\text{eff}}&=&\mu\vec{B}\cdot\vec{\sigma}-\frac12i\hbar\kappa\sigmap\sigmam\nonumber\\
&=&\mu B\left(
     \begin{array}{cc}
       \cos\theta-\frac12ig & \sin\theta e^{-{2\pi zi}/{L}} \\
       \sin\theta e^{{2\pi zi}/{L}} & -\cos\theta \\
     \end{array}
   \right),
\end{eqnarray}
where $g=\kappa\hbar/\mu B$ is dimensionless dissipation rate. For
each fixed $z$, this non-Hermitian Hamiltonian has two right
eigenstates
\begin{eqnarray*}
&&\ket{\psi^R_+}=\frac1{N}\left(\cos\frac{\alpha}2\1+\sin\frac{\alpha}2e^{{2\pi
zi}/L}\0\right)\nonumber\\
&&\ket{\psi^R_-}=\frac1{N}\left(\sin\frac{\alpha}2\1-\cos\frac{\alpha}2e^{{2\pi
zi}/L}\0\right),
\end{eqnarray*}
two left eigenstates
\begin{eqnarray*}
&&\bra{\psi^L_+}={N}\left(\cos\frac{\alpha}2\bra{1}+\sin\frac{\alpha}2e^{{-2\pi
zi}/L}\bra{0}\right)\nonumber\\
&&\bra{\psi^L_-}={N}\left(\sin\frac{\alpha}2\bra{1}-\cos\frac{\alpha}2e^{{-2\pi
zi}/L}\bra{0}\right),
\end{eqnarray*}
and corresponding eigenvalues (in units of $\mu B$)
\begin{eqnarray*}
E_{\pm}=-\frac12ig\pm\frac12\sqrt{16-g^2-8ig\cos\theta}
\end{eqnarray*}
In the above expressions, the angle $\alpha$ is defined as
\begin{eqnarray*}
\tan\alpha=\frac{4\sin\theta}{4\cos\theta-ig}
\end{eqnarray*}
and the normalized coefficient $N$ is
\begin{eqnarray*}
N=\sqrt{\left|\cos\frac{\alpha}2\right|^2+\left|\sin\frac{\alpha}2\right|^2}.
\end{eqnarray*}
Note that for a non-zero dimensionless dissipation rate $g$,
$\alpha$ is a complex number. In this case, the relation
$\sin^2\frac{\alpha}2+\cos^2\frac{\alpha}2=1$ holds while
$\left|\sin\frac{\alpha}2\right|^2+\left|\cos\frac{\alpha}2\right|^2=1$
does not. Put these eigenstates and eigenvalues into
Eq.(\ref{slowzeroeqn}), we obtain the zero-order Hamiltonian for the
spatial variables as
\begin{eqnarray}
&&\mathcal{H}_n=-\frac{\hbar^2}{2M}(\nabla-i\vec{A}_n)^2+E_n,\nonumber\\
&&\vec{A}_n=i\bra{\psi_n^L}\nabla\ket{\psi_n^R},~~~~~n=+,-.
\end{eqnarray}
With these knowledge, we study the population transfer among the
internal state for the quantum system. Suppose that we prepare the
spin of the neutron in the state $\ket{{+}\frac12}$ initially and
manipulate the particle moving from $z=0$ to $z=L$ in a fixed time
interval $T=3$ (in units of $\pi\hbar/\mu B$). Setting
$\theta=\frac{\pi}4$, we study the polarization of the neutron along
$z$ axis versus time $t$ with different dimensionless dissipation
rate $g$, the results are shown  in Fig.\ref{FIG:example2_1}. In the
simulation we take $\mathcal{N}=400$ trajectories. Some feature can
be seen from the figure: When the dissipation is absent,  the
polarization along $z$ axis oscillates between 0 and 1 as a cosine
function of time. The dissipation leads the polarization damping to
$-1$ oscillatingly. The stronger the dissipation is, the faster of
the damping is. To measure the validity condition, we define a
function
\begin{eqnarray}
\Gamma(g)=\max\left\{\left|\frac{\bra{\varphi_{n',k'}^{L[0]}}O_{n',n}\ket{\varphi_{n,k}^{R[0]}}}
{E_{n',k'}^{[0]}-E_{n,k}^{[0]}}\right|\right\},\label{Gammag}
\end{eqnarray}
 where
$O_{n',n}=-\frac{\hbar}{2M}(2\bra{\psi_{n'}^L}\nabla\ket{\psi_n^R}\nabla+
\bra{\psi_{n'}^L}\nabla^2\ket{\psi_n^R}),$
 to characterize the
violation of the BO condition. From Fig.\ref{FIG:example2_2} we can
see that the spin relaxation benefits the approximation. This is
same as that in our previous work \cite{Huang2009PRA,Yi2007JPB}, and
it can also be understood as that the dissipation in fast variables
benefit the approximation, because it accelerates the moving of fast
variables and the difference between the two types of variables
becomes more evident.

It is time to give a detailed comparison between this method and our
previous approach \cite{Huang2009PRA}. We  note that the differences
come from the two methods itself. This leads to the following
distinct features: (1) In Ref. \cite{Huang2009PRA}, the extension is
done by effective Hamiltonian approach, which requires to extend the
Hilbert space. In the present paper, the extension is done according
to the quantum trajectory approach that does not require to extend
the Hilbert space. In addition, in the present paper,  if the
initial state is pure, the state will always pure in the evolution.
(2) In the effective Hamiltonian approach, the extension is simple
and straightforward, however, the eigenstates of the Hamiltonian
including  the fast variables  may  not be  physical states,
although it gives a correct dynamics. For example, the last three
eigenstates of $H_S^{\text{T}}$ in Ref. \cite{Huang2009PRA} are not
physical states. In quantum trajectory approach, the eigenstates are
all physical states. (3) The complexity is different.  The
analytical solution for the non-jump trajectory is easier than the
effective Hamiltonian solution. For example, the last three
eigenstates for the fast variables $H_S^{\text{T}}$
\cite{Huang2009PRA} is a cubic equation, whose solution is
complicated. For a high dimensional open system, the problem become
more complicated. But it is relatively easy to solve the eigenstates
in this paper, i.e., the non-Hermitian Hamiltonian in non-jump
trajectory can be solved more easily than the effective Hamiltonian
in Ref. \cite{Huang2009PRA}. (4) The method in present paper  is
more accurate to treat the jump term in the master equation. This
can be understood as follows. When the dissipation in slow variables
is considered, the non-diagonal elements of perturbation
$\mathcal{H}_{\mathcal{P}}$ can be divided into two parts as
$H_{n,m}=\bra{\psi_n}H_s\ket{\psi_m}-\frac12i\sum_k\bra{\psi_n}X_k^{\dag}
X_k\ket{\psi_m}$. The first part is the same as that in closed
system while the second part only contains a term of dissipation.
Other part of dissipation is recovered via the jump process.
Compared with effective Hamiltonian solution, in which the
perturbation includes all parts of dissipation, the quantum
trajectory solution is more exact. Before closing this paper,  we
emphasize that all the discussions in both two methods are
restricted to the non-degenerate energy level, i.e., we assume the
closed system Hamiltonian is non-degenerate. However, when the
dissipation is taken into account, new degeneracy  is introduced
\cite{Tay2007PRA} in both methods. For example, in the second
example in this paper, even the original system is non-degenerate,
the non-Hermitian Hamiltonian $H_{\text{eff}}$ can be degenerate at
$\theta=\frac{\pi}2$ and $g=4$. In Ref. \cite{Huang2009PRA} the
degenerate occurs at $\theta=\frac{\pi}2,~z=z^{\text{A}}$ and $g=8$.
Obviously,  the degenerate points in both methods are different, so
these two methods are  complementary in this sense, in other words,
when one method is not available due to the degeneracy, we can
choose another method.

In summary, we have extended the BO approximation from closed system
to open system by the quantum trajectory approach. An assumption
that the dissipation is in Lindblad form is required. The BO
approximation is used in the non-jump trajectory and the dynamics
can be recovered by the Monte Carlo wave function method. As
illustrations, we give two examples to detail our method. The
results show that our method can reproduce the dissipation dynamics
for such systems efficiently. A detailed comparison with our
previous work is also given and discussed.


\ \ \\
This work is supported by NSF of China under Grant Nos. 10775023,
10935010 and 10905007.
\\

\end{document}